\documentclass[aps,prb,reprint]{revtex4-1}
\usepackage{amsmath,amssymb}
\usepackage{txfonts}
\usepackage[dvipdfmx]{graphicx}
\usepackage{color}
\usepackage{here}

\begin{document}

\title{Analysis of Kohn-Sham Eigenfunctions Using a Convolutional Neural Network
in Simulations of the Metal-insulator Transition in Doped Semiconductors }

\author{Yosuke Harashima$^{1,4}$, Tomohiro Mano$^2$, Keith Slevin$^3$, and Tomi Ohtsuki$^2$}
\affiliation{
  $^1$Institute of Materials and Systems for Sustainability,
  Nagoya University,
  Nagoya,
  Aichi 464-8601,
  Japan
  \\
  $^2$Physics Division,
  Sophia University,
  Chiyoda,
  Tokyo 102-8554,
  Japan
  \\
  $^3$Department of Physics,
  Osaka University,
  Toyonaka,
  Osaka 560-0043,
  Japan
  \\
  $^4$Center for Computational Sciences,
  University of Tsukuba,
  Tsukuba,
  Ibaraki 305-8577,
  Japan}

\begin{abstract}
  Machine learning has recently been applied to many problems in condensed matter physics.
  A common point of many proposals is to save computational cost by training the machine with data from a simple example
  and then using the machine to make predictions for a more complicated example.
  Convolutional neural networks (CNN), which are one of the tools of machine learning, have proved to work well for assessing eigenfunctions in disordered systems.
  Here we apply a CNN to assess Kohn-Sham eigenfunctions obtained in density functional theory (DFT)
  simulations of the metal-insulator transition of a doped semiconductor.
  We demonstrate that a CNN that has been trained using eigenfunctions from a simulation of a doped semiconductor that neglects electron spin
  successfully predicts the critical concentration when presented with eigenfunctions from simulations that include spin.
\end{abstract}

\maketitle

\section{Introduction}

Machine learning has proven to be very useful in condensed matter physics.\cite{Mehta19,Carleo19,Ohtsuki20}
It has been applied to the classification of phases in spin systems\cite{Carrasquilla17,Nieuwenburg17},
interacting systems\cite{Broecker17} and disordered systems\cite{Tomoki16} as well as topological systems.
\cite{Tomoki17,Zhang17a,Zhang17b,Yoshioka18,Araki19,Mano19}
There are also suggestions to use machine learning for
calculating atomistic potentials\cite{BePa2007,Takahashi17,Wenwen17,Bartok18,Babaei19,Byggmastar19}
and performing materials search.\cite{Takahashi16,Yamashita18,FuHaHoMi2019,Harashima20}

Recent work on the Anderson transition in three dimensions has shown that a CNN can detect the transition point
from the spatial profile of the eigenfunction intensity.\cite{Tomoki17,MaOh2017}
Though the precision of the estimate of the critical point was less than that achieved with finite size scaling (FSS), 
the CNN predicted the critical point from the simulation of only a single system size.
It also showed generalisation capability; once trained for Anderson's model of localisation\cite{Anderson58},
it was successfully applied to quantum percolation without further training.\cite{MaOh2017,Ohtsuki20}
This suggests that a suitably trained CNN might also be usefully applied to study other transitions, 
for example, the metal-insulator transition in doped semiconductors.

A metal-insulator transition as a function of doping concentration is observed in numerous semiconductors
when they are doped with impurities.\cite{RoAnThBh1980,StHoLaMaLo1993,Lo2011,ItWaOoHaOh2004}
This transition is thought to be a zero temperature continuous quantum phase transition
in which both disorder due to the random positions of the impurities in the semiconductor and
interactions between electrons play important roles.

Recently there have been attempts to better understand this transition, and, in particular, its critical phenomena, by
studying the FSS of multifractal measures calculated from
eigenfunctions obtained by using DFT simulations of doped semiconductors.\cite{HaSl2012,HaSl2014,CaHiRo2019,Carnio19}
In these studies, it has been found that the eigenfunction at the Fermi level is Anderson localised at 
low doping concentration but becomes delocalised when the doping concentration is sufficiently high.
Supposing this coincides with the metal-insulator transition in the doped
semiconductor, this provides an estimate of the critical concentration.

In Refs. \onlinecite{HaSl2012, HaSl2014} the role of electron spin was ignored.
However, the true spin configuration is expected to be paramagnetic with the local magnetic moment randomly distributed.
Unfortunately, DFT calculations which include the electron spin are considerably more time-consuming than calculations for spinless electrons.
It would save computational time, if data from simulations with spinless electrons could be used to train the CNN.
However, this would only be useful if such a CNN has the necessary generalisation capability.

In this paper, we demonstrate that a CNN that has been trained using Kohn-Sham eigenfunctions for spinless electrons
successfully predicts the critical concentration when presented with Kohn-Sham eigenfunctions obtained in calculations that include spin.
We also use the same CNN to analyse Kohn-Sham eigenfunctions 
obtained from simulations with spinless electrons for compensated semiconductors.

\section{Models and Methods}

\begin{figure*}[th]
  \includegraphics[width=0.30\hsize]{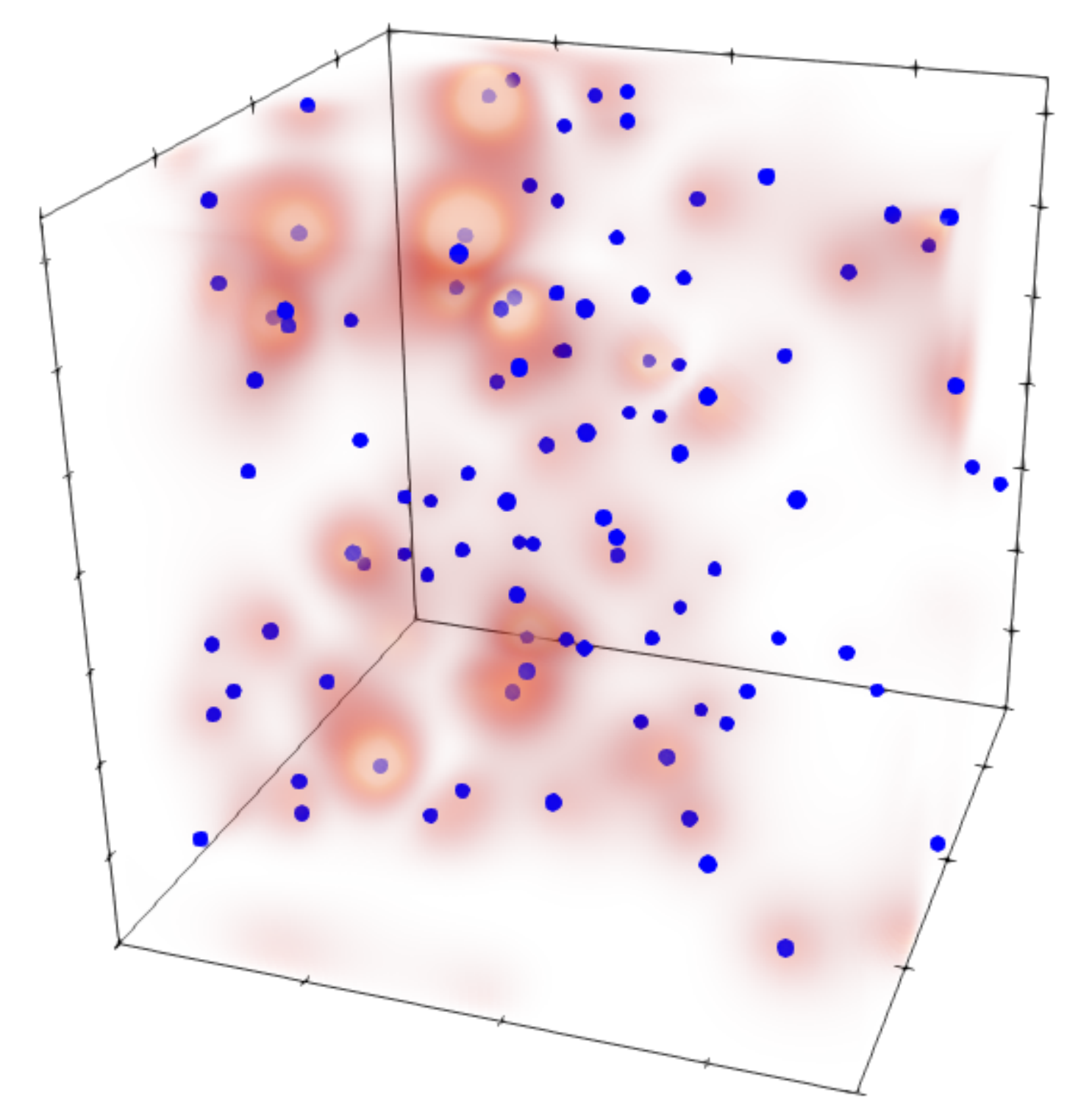}
  \includegraphics[width=0.30\hsize]{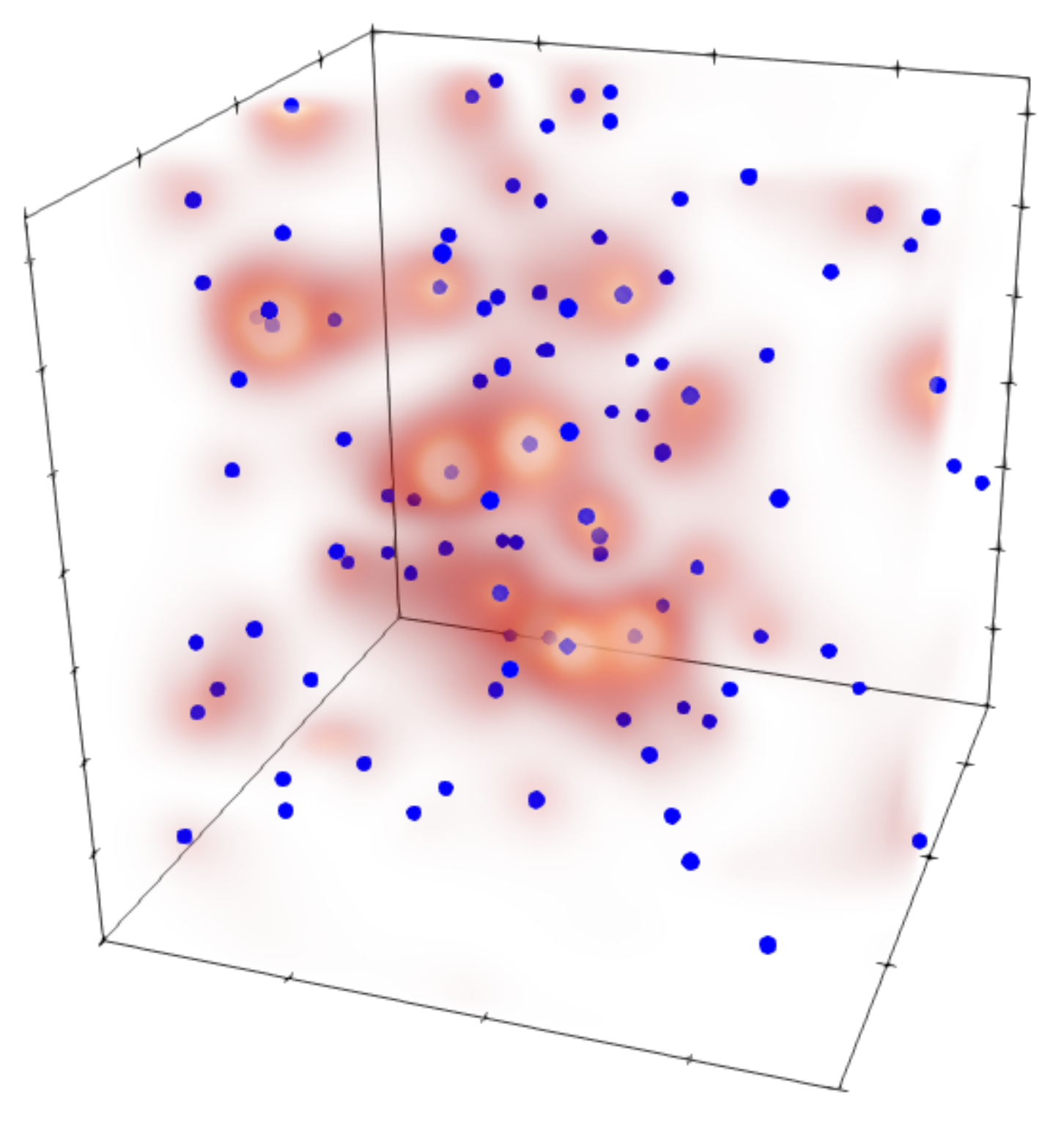}
  \includegraphics[width=0.32\hsize]{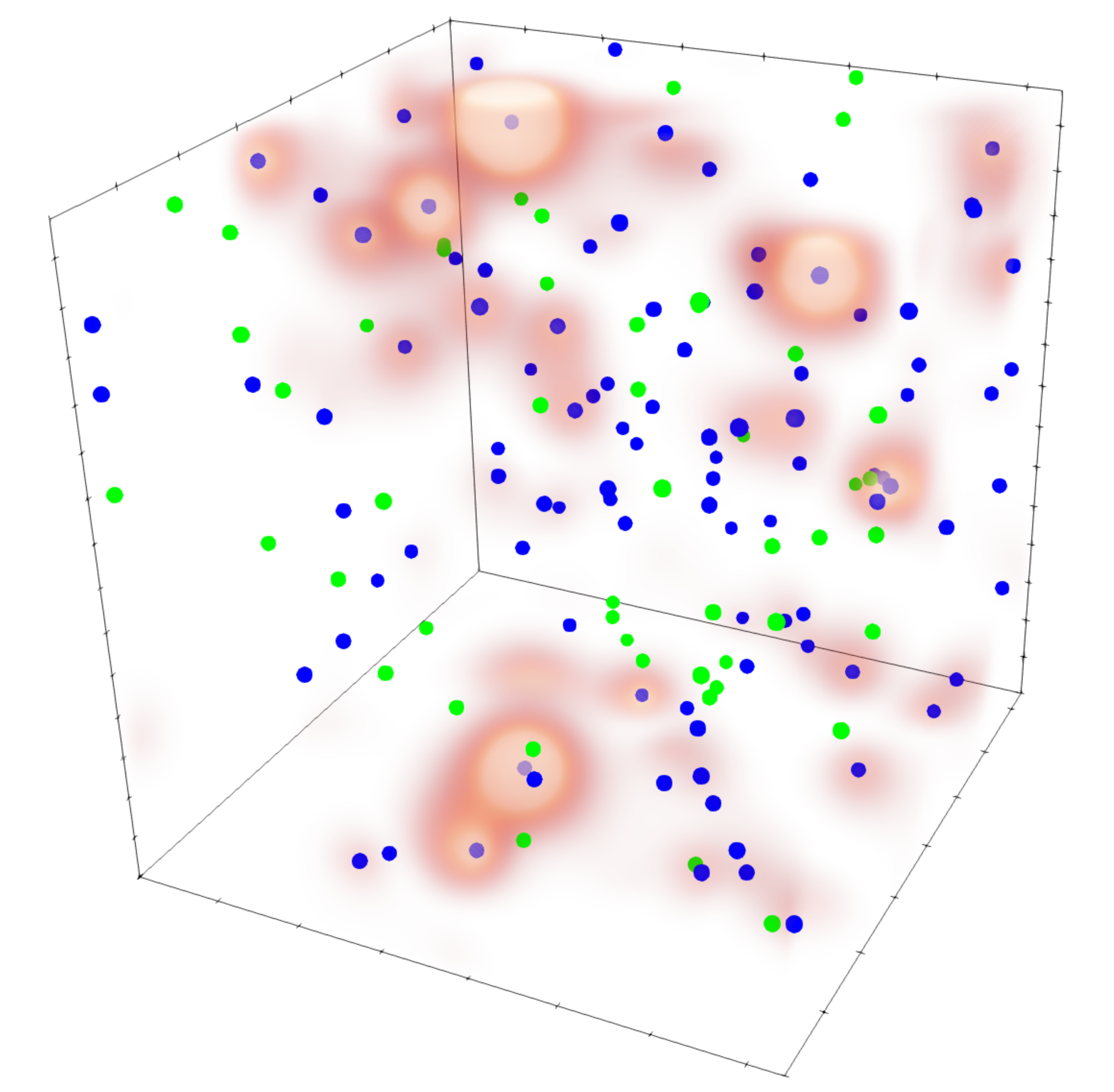}  
  \caption{ Example of impurity distributions and Kohn-Sham eigenfunctions.
    Blue and green dots are donor and acceptor ions, respectively.
    Red shading indicates the square of the highest occupied Kohn-Sham eigenfunction. On the left a spin-up eigenfunction, and in the centre a spin-down eigenfunction.
    On the right a Kohn-Sham eigenfunction for a spinless compensated sample.}
  \label{fig1}
\end{figure*}

The metal-insulator transition in a doped-semiconductor can be studied theoretically,
at a certain level of approximation,
using a model in which electrons with an effective mass $m^*$ move in an effective medium
with relative dielectric constant $\varepsilon_{\mathrm{r}}$.
This leads to the consideration of the Hamiltonian
\begin{align}
  \mathcal{H} =& -\dfrac{1}{2m^{*}} \sum_{i} \nabla_{i}^{2}
  - \dfrac{1}{\varepsilon_{\mathrm{r}}} \sum_{i,I} \dfrac{Z_{I}}{|\Vec{r}_{i}-\Vec{R}_{I}|}
  \nonumber
  \\
  &+ \dfrac{1}{2\varepsilon_{\mathrm{r}}} \sum_{i \neq j} \dfrac{1}{|\Vec{r}_{i}-\Vec{r}_{j}|}
  + \dfrac{1}{2\varepsilon_{\mathrm{r}}} \sum_{I \neq J} \dfrac{Z_{I}Z_{J}}{|\Vec{R}_{I}-\Vec{R}_{J}|}.
  \label{eq:hamiltonian}
\end{align}
Here, $\Vec{r}_{i}$ and $\Vec{R}_{I}$ are the positions of the electrons and the impurity ions, respectively,
$Z_{I}$ is the ionic charge value, which is $+1$ for a donor and $-1$ for an acceptor, and
we use Hartree atomic units.
We consider an ensemble of three dimensional cubic systems of linear size $L$.
In each system, $N_{\mathrm{D}}$ donor impurity ions and $N_{\mathrm{A}}$ acceptor impurity ions
are randomly distributed (see Fig.~\ref{fig1}).
The corresponding concentrations are
\begin{equation}
    n_{\mathrm{D}} =  N_{\mathrm{D}}/V,
\end{equation}
and
\begin{equation}
     n_{\mathrm{A}} = N_{\mathrm{A}}/V,
\end{equation}
with
\begin{equation}
    V=L^3
\end{equation}
being the volume of the system.

\begin{table*}[t]
  \caption{Summary of results. 
    Spin degree of freedom, 
    compensation (ratio), 
    critical concentrations ($n_{\mathrm{c}}^{\mathrm{CNN}}$ for CNN and $n_{\mathrm{c}}^{\mathrm{MFA}}$ for MFA).
    The model (A), (B), and (C) are discussed in Sec.~3.1, 3.2, and 3.3 in the manuscript, respectively.
    The estimated fluctuations and mean values of the effective Hamiltonian $\sigma_{\epsilon}$, $\mu_{t}$, and $\sigma_{t}$ are also shown.
    We discuss these values in Sec.~3.4.}
  \vspace{5pt}
  \begin{center}
  \begin{tabular}{cccccccc}
    \hline \hline
    \\[-8pt]
    & spin & compensation & $n_{\mathrm{c}}^{\mathrm{CNN}}$ & $n_{\mathrm{c}}^{\mathrm{MFA}}$ & $\sigma_{\epsilon}$ & $\mu_{t}$ & $\sigma_{t}$ \\[2pt]
    &      &              & $10^{18} \ \mathrm{cm}^{-3}$ & $10^{18} \ \mathrm{cm}^{-3}$ & eV & eV & eV \\
    \hline
    (A) & $\times$   & $\times$   & 1.09 & 1.09 & 0.0068 & $-$0.0138 & 0.0105 \\
    (B) & \checkmark & $\times$   & 1.56 & 1.55 & 0.0054 & $-$0.0103 & 0.0067 \\
    (C) & $\times$   & 0.5        & 1.80 &      & 0.0135 & $-$0.0157 & 0.0120 \\
    \hline \hline
    \end{tabular}
  \end{center}
  \label{table:result}
\end{table*}

As in previous studies,\cite{HaSl2012,HaSl2014}
for a given configuration of the impurities, we attempt to find the ground state of Eq.~\eqref{eq:hamiltonian}
for 
\begin{equation}
    N =  N_{\mathrm{D}} - N_{\mathrm{A}}
\end{equation}
electrons,
using the Kohn-Sham formulation of DFT.\cite{HoKo1964,KoSh1965}
This involves finding the self-consistent solutions of the Kohn-Sham equations,
\begin{equation}\label{eq:kohn-sham_1}
  \left(-\dfrac{1}{2m^{*}}\nabla^{2} + V^{\sigma}_{\mathrm{eff}}\right) \psi_{i}^{\sigma}(\Vec{r})
  = \epsilon^{\sigma}_{i}\psi^{\sigma}_{i}(\Vec{r}).
\end{equation}
Here, $\psi^{\sigma}_i$ are the Kohn-Sham eigenfunctions, $\epsilon^{\sigma}_i$ the Kohn-Sham eigenvalues, and $\sigma$ is the spin.
The electron density of the ground state $n \left( \vec{r} \right)$ is then the sum of the spin resolved electron densities
\begin{equation}
  n \left( \vec{r} \right) = n^{\uparrow} \left( \vec{r} \right) + n^{\downarrow} \left( \vec{r} \right),
\end{equation}
where 
\begin{equation}
    n^{\sigma} \left( \vec{r} \right) = \sum_{i:\text{occupied}} \left| \psi^{\sigma}_{i} \right|^2.
\end{equation}
The effective potentials $V^{\sigma}_{\mathrm{eff}}$ are the sum of three terms
\begin{equation}\label{eq:kohn-sham_2}
  V_{\mathrm{eff}}^{\sigma} = V_{\mathrm{ext}} + V_{\mathrm{Hartree}} + V^{\sigma}_{\mathrm{XC}}.
\end{equation}
The first term is the external potential due to the impurity ions.
The second term is the Hartree potential of the electrons.
The third term $V_{\mathrm{XC}}^{\sigma}$ is the exchange-correlation potential
\begin{equation}
  V^{\sigma}_{\mathrm{XC}} = \frac{\delta E_{\mathrm{XC}} \left[n^{\uparrow}, n^{\downarrow} \right]}{\delta n^{\sigma}}.
\end{equation}
Here, $E_{\mathrm{XC}}$ is the exchange-correlation energy, which is a functional of the spin up and spin down electron densities,
or equivalently the electron density $n \left( \vec{r} \right)$ and the spin density
\begin{equation}
  \zeta \left( \vec{r} \right) = \frac{n^{\uparrow} \left( \vec{r} \right) - n^{\downarrow} \left( \vec{r} \right)}{n \left( \vec{r} \right)}.
\end{equation}
We use the local density approximation\cite{GuLuWi1974,JaMoWi1975}
\begin{equation}
  E_{\mathrm{XC}} \approx E_{\mathrm{XC}}^{\mathrm{LDA}} =
  \int d^3r \; \epsilon_{\mathrm{XC}} \left( n \left( \vec{r} \right), \zeta \left( \vec{r} \right) \right) n \left( \vec{r} \right),
\end{equation}
with the form of $\epsilon_{\mathrm{XC}}$ given in Refs.~\onlinecite{GuLuWi1974,JaMoWi1975}.
Solving these equations self-consistently, we obtain the eigenfunctions $\psi^{\sigma}_{i}$.
We focus on the highest occupied Kohn-Sham eigenfunction, i.e. the occupied eigenstate with the largest eigenvalue.
For brevity in what follows we denote this eigenfunction simply as $\psi$.

We train the CNN so that it can correctly determine the localised and delocalised phases
from the eigenfunction.
The input is the intensity $|\psi |^2$  and
the output is the probability $p_\mathrm{loc}$
that the eigenfunction is in the localised phase.\cite{Ohtsuki20}
We perform supervised training, i.e,
we prepare a correctly labelled data set (training data) in advance to optimise the weight parameters of the CNN.
The hyper-parameters of the network structure are similar to the ones used in Refs.~\onlinecite{MaOh2017,Ohtsuki20}
for Anderson's model of localisation in three dimensions.
Training data is prepared by simulating a system of spinless electrons without acceptors, 
i.e., by solving the Kohn-Sham equations subject to the constraint of complete spin-polarisation
\begin{equation}\label{eq:spin_polarized}
  \zeta \left( \vec{r} \right) = 1
\end{equation}
everywhere and $N_{\mathrm{A}}=0$.

To check the ability of the CNN to generalise, i.e., to determine the correct phases when presented with an unlabelled data set of eigenfunctions
for a system with spin or compensation, we need an independent estimate of the critical concentration.
For this purpose we use multi-fractal finite size scaling.
This method was applied to the Anderson transition in three dimensions in Refs.~\onlinecite{RoVaSlRo2010,RoVaSlRo2011}
and subsequently to the Anderson transitions in three dimensions in all the Wigner-Dyson classes\cite{Ujfalusi15, Lindinger17} 
as well as quantum percolation\cite{Ujfalusi14}.

In multi-fractal finite size scaling the system size dependence of the effective multifractal exponent $\tilde{\alpha}_{0}$ is analysed.
This exponent is defined as,
\begin{equation}
  \tilde{\alpha}_{0} \equiv \dfrac{\lambda^{3} \langle S_{0} \rangle}{\ln \lambda},
  \label{eq:mfe_1}
\end{equation}
where $S_{0}$ and $\lambda$ are defined as follows.
Calculation of Eq.~\eqref{eq:mfe_1} involves coarse-grained eigenfunction intensities.
A three dimensional cubic system of linear size $L$ is divided into boxes (indexed by the label $k$) of linear size $l$
and the eigenfunction intensities integrated over each box
\begin{equation}
 \mu_{k} \equiv \int_{k} d^{3}r \left|\psi(\Vec{r})\right|^{2}.
  \label{eq:mfe_3}
\end{equation}
The ratio of the box size to the system size is denoted by
\begin{equation}
  \lambda \equiv \dfrac{l}{L}.
  \label{eq:mfe_4}
\end{equation}
The quantity $S_0$ is obtained by summing over all the boxes as follows
\begin{equation}
  S_{0} \equiv \sum_{k} \ln \mu_{k}.
  \label{eq:mfe_2}
\end{equation}
The angular brackets $\langle\cdots \rangle$ denotes an ensemble average.

As discussed in Refs.~\onlinecite{RoVaSlRo2010,RoVaSlRo2011}, for a system with dimensionality $d$ (here $d=3$) and with $\lambda$ held fixed, 
we expect $\tilde{\alpha}_{0} \rightarrow \infty$ as $L\rightarrow \infty$ in the localised or insulating phase,
while $\tilde{\alpha}_{0} \rightarrow d$ as $L\rightarrow \infty$ in the delocalised or metallic phase.
At the critical point, the system size dependence of $\tilde{\alpha}_0$ disappears.
We can, therefore, find the critical point as a crossing between curves for systems of two different sizes $L$ with the box size adjusted such that
the value of $\lambda$ is the same for both curves (for example, see Fig.~\ref{fig4}).

It should be noted that the true multifractal exponent $\alpha_0$ is obtained only in the limit that $\lambda\rightarrow0$ and for this reason we use the adjective ``effective" (and a tilde) above.

\section{Results and Discussion}

For ease of comparison with the well studied case of Si, in what follows we set $m^{*} = 0.32 m_{\mathrm{e}}$ and $\varepsilon_{\mathrm{r}} = 12.0$,
which are the appropriate values for electrons in Si.
However, for Eq. (\ref{eq:hamiltonian}), this amounts only to a re-scaling of the units and does not affect the analysis in any fundamental way.
We discuss three models: 
the spinless uncompensated model (A) in Sec.~3.1, 
the uncompensated model with spin (B) in Sec.~3.2, and 
the spinless compensated model (C) in Sec.~3.3.
The estimated critical concentrations are summarised in Table~\ref{table:result}.

\subsection{Spinless electron model of doped semiconductor}

\begin{figure}[t]
  \includegraphics[width=\hsize]{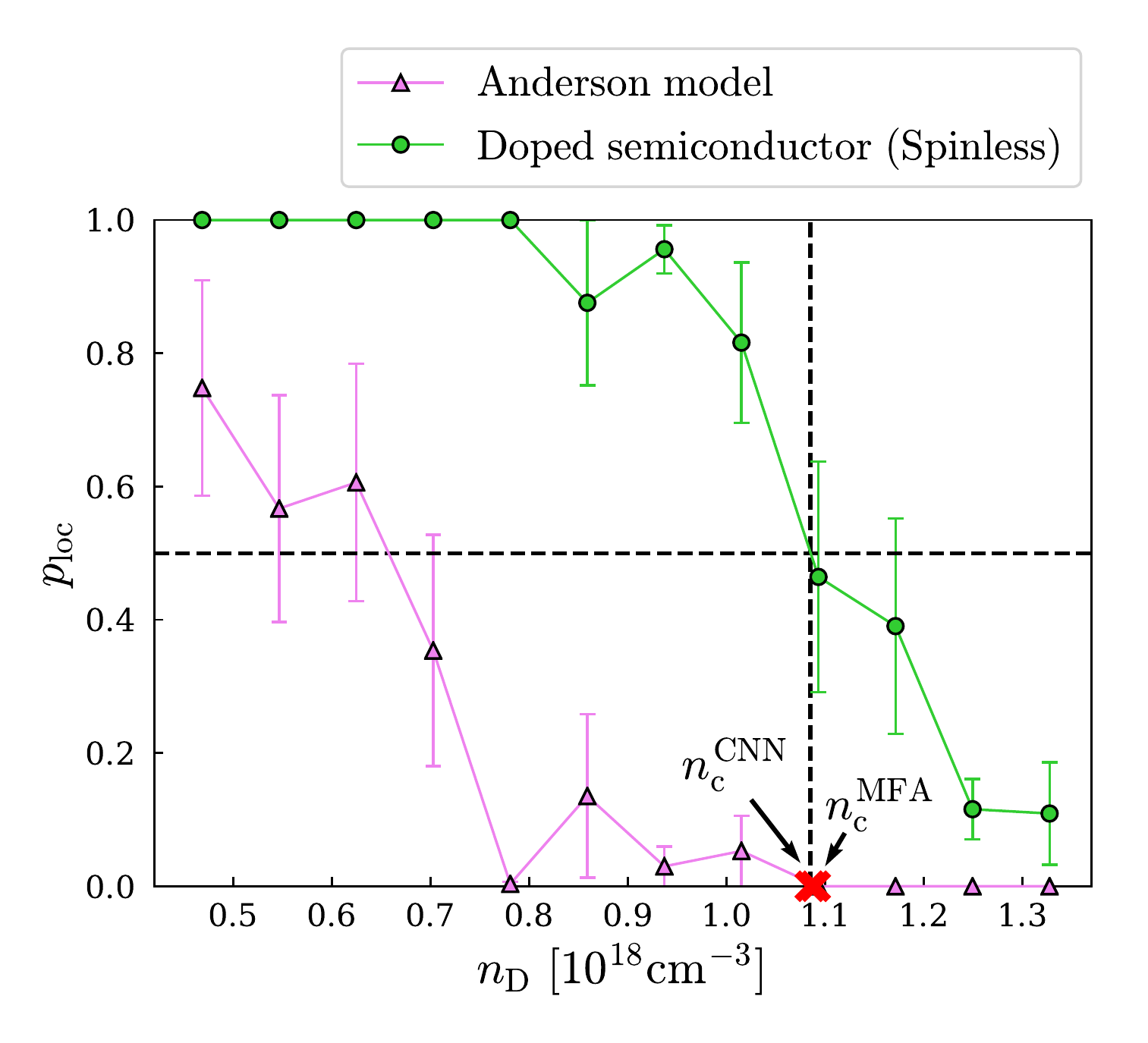}
  \caption{
    The probability $p_{\mathrm{loc}}$ that a Kohn-Sham eigenfunction is localised as a function of the
    doping concentration $n_{\mathrm{D}}$.
    We compare the probabilities reported by two CNNs: one trained with
    data for Anderson's model of localisation, and the other with data for a spinless uncompensated doped semiconductor.
    The system size is $L\approx 400.1\mathrm{\AA}$ and the probabilities have been averaged over 8 samples.
    The prediction for the critical concentration is taken as the concentration where $p_{\mathrm{loc}}=0.5$.
    For guides to the eye, dashed lines are drawn.
    The CNN trained with data for a spinless uncompensated doped semiconductor predicts a critical concentration
    $n_{\mathrm{c}}^{\mathrm{CNN}}\approx 1.09\times 10^{18}\mathrm{cm}^{-3}$.
    The critical concentration of multifractal analysis 
    (MFA)~\cite{HaSl2014} $n_{\mathrm{c}}^{\mathrm{MFA}}\approx 1.09\times 10^{18}\mathrm{cm}^{-3}$ is also shown for comparison.}
  \label{fig2}
\end{figure}

We consider two ways of training the CNN.
The first is with a labelled data set of eigenfunctions of Anderson's model of localisation.
The second is with a labelled data set of Kohn-Sham eigenfunctions for a spinless model of a doped semiconductor.

For Anderson's model of localisation the training set and the CNN structure are the same as in Ref.~\onlinecite{MaOh2017} 
except the input system size, which is $42\times 42\times 42$ in the present case.
For the doped semiconductor the training set consists of the highest occupied Kohn-Sham eigenfunctions obtained in
simulations of 1,000 samples each for doping concentrations of $n_{\mathrm{D}} \approx$ 0.86 $\times$ 10$^{18}$ cm$^{-3}$, which is in the insulating regime,
and $n_{\mathrm{D}} \approx$ 1.33 $\times$ 10$^{18}$ cm$^{-3}$, which is in the metallic regime, and system size $L \approx 400.1$\AA.

After training the CNNs are then presented with an unlabelled set of Kohn-Sham eigenfunctions for the same spinless model of a doped semiconductor.
The results are shown in Fig.\ref{fig2}.
The critical concentration obtained in a previous study is shown by the red cross.
As can be seen immediately, a CNN trained with Anderson's model of localisation
fails to predict the correct critical concentration for the semiconductor.
The CNN trained with the model of the doped semiconductor naturally gives the correct critical concentration.

\subsection{Model of doped semiconductor including electron spin}

\begin{figure}[t]
  \includegraphics[width=0.985\hsize]{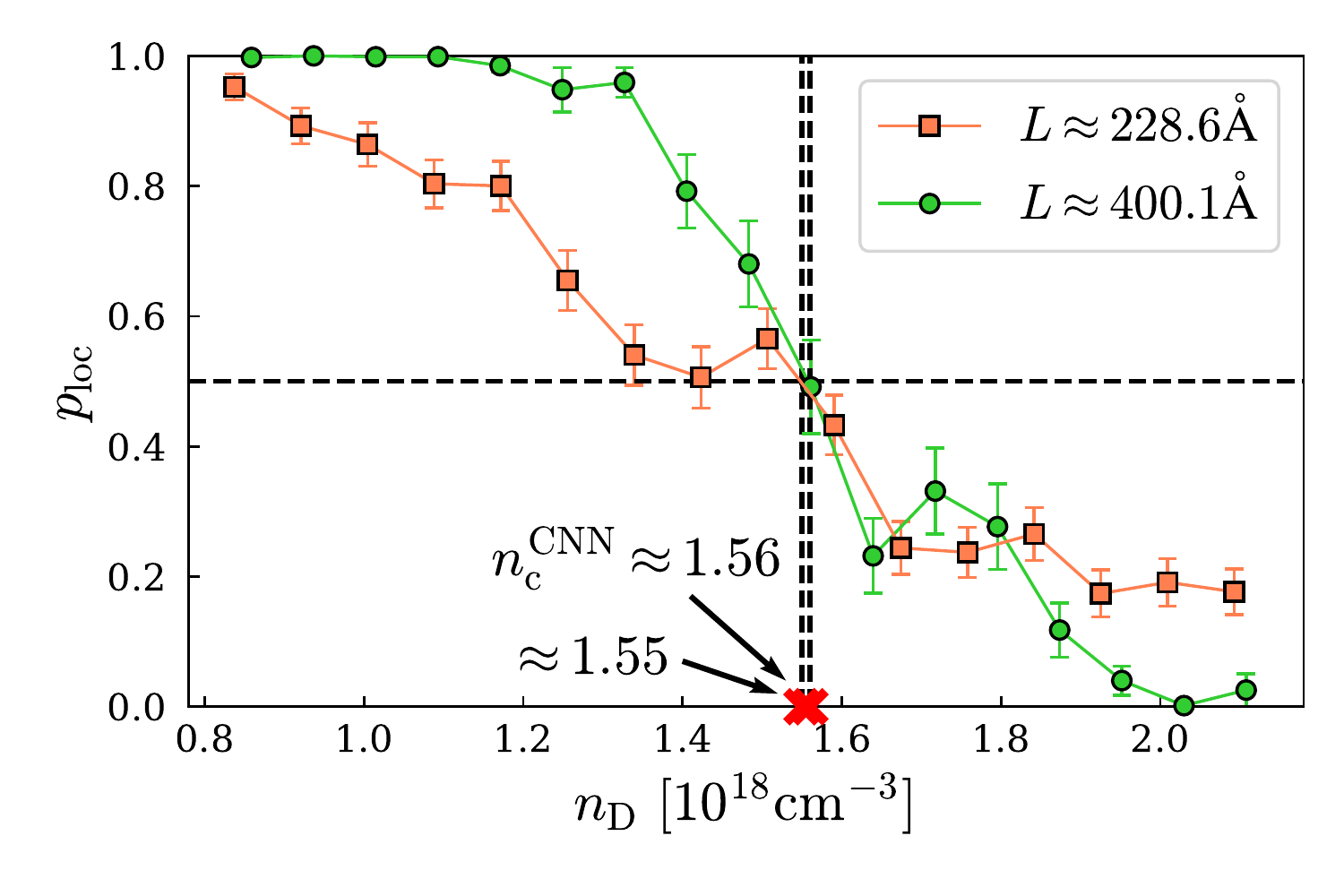}
  \caption{The probability $p_{\mathrm{loc}}$ that a Kohn-Sham eigenfunction is localised as a function of
  the doping concentration $n_{\mathrm{D}}$ for a system with spin.
  The CNN used has been trained with data for the spinless uncompensated doped semiconductor.
  The critical concentration is estimated as the concentration where $p_{\mathrm{loc}}=0.5$.
  The estimated values are $n_\mathrm{c}\approx 1.56\times 10^{18}\mathrm{cm}^{-3}$ for $L\approx 400.1 \mathrm{\AA}$ and $n_\mathrm{c}\approx 1.55\times 10^{18}\mathrm{cm}^{-3}$ for $L\approx 228.6 \mathrm{\AA}$.}
  \label{fig3}
\end{figure}

\begin{figure}[t]
  \includegraphics[width=\hsize]{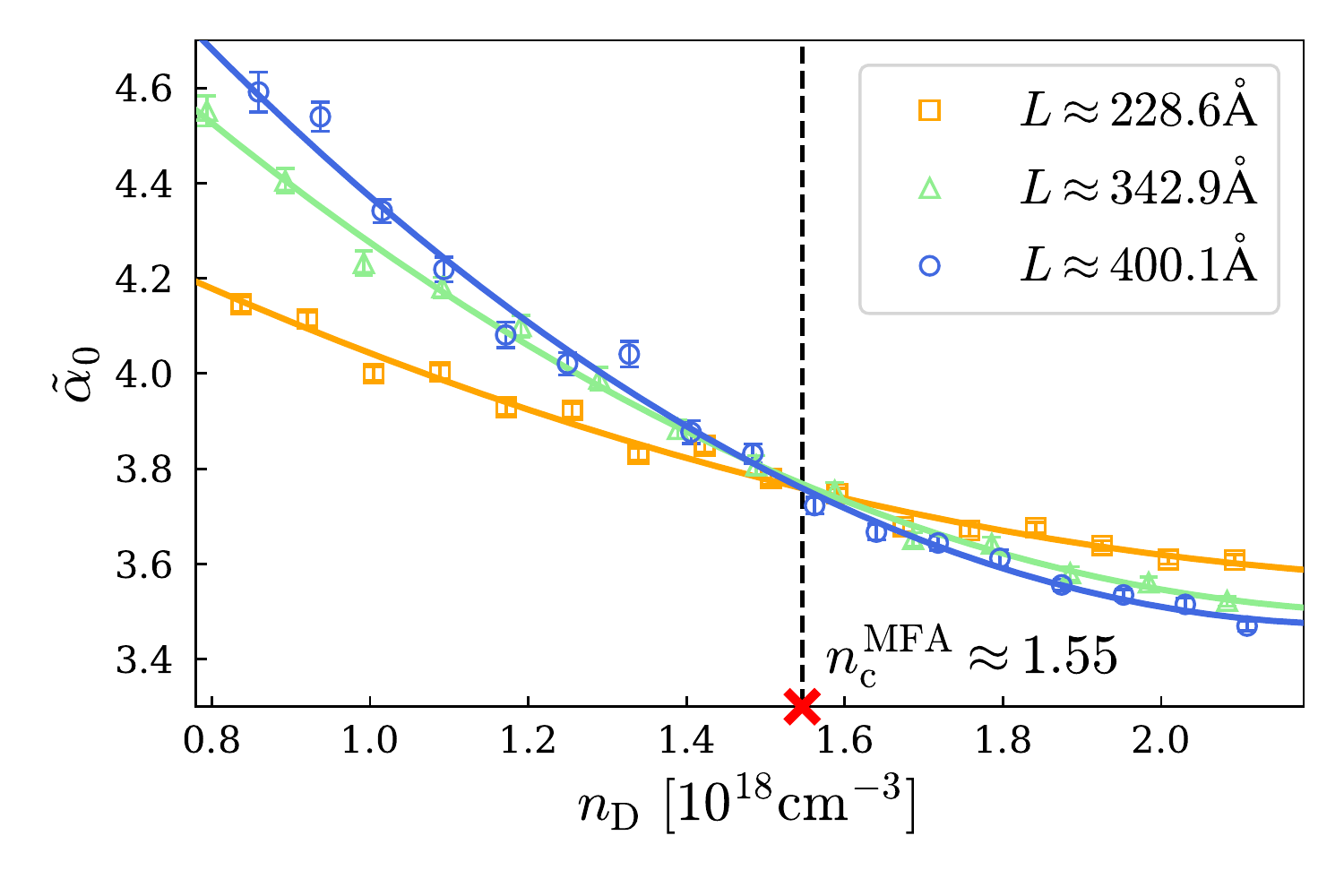}
  \caption{ 
  Estimation of the critical concentration for a system with spin using multifractal analysis.
  Three system sizes were simulated: $L\approx228.6$\AA (square), $L\approx342.9$\AA (triangle), and $L\approx400.1$\AA (circle), with  
  ensembles of 30, 30, and 20 samples, respectively.
  The crossing point of the curves for the largest and smallest systems is taken as the estimate of the critical concentration.
  The value found $n_\mathrm{c}\approx 1.55\times 10^{18}\mathrm{cm}^{-3}$ is in good
  agreement with the value found using the CNN (see Fig. \ref{fig3}).}
  \label{fig4}
\end{figure}

We use a CNN, trained as before with a labelled data set for a spinless model of a doped semiconductor,
to assess Kohn-Sham eigenfunctions obtained from a model of a doped semiconductor that includes spin.
That is in a model where the condition of complete spin polarisation Eq.~(\ref{eq:spin_polarized}) is removed.
Since the spin configuration must also be optimised more iterations are required to find the self-consistent solutions
of the Kohn-Sham equations.
When presented with the resulting eigenfunctions the CNN reports a probability that an eigenfunction is localised.
We average this over the highest occupied spin-up and spin-down eigenfunctions and denote the result $p_\mathrm{loc}$.

The results obtained after averaging $p_\mathrm{loc}$ over an ensemble of 20 samples with system
size $L \approx 400.1 $\AA\;  are shown in Fig.~\ref{fig3}.
The concentration corresponding to  $p_\mathrm{loc}=0.5$ is about $n_\mathrm{c}\approx 1.56\times$ 10$^{18}$ cm$^{-3}$.
We also plot in the figure the probability for $L\approx 228.6 \mathrm{\AA}$, where an average over 50 samples is taken.
Here, this CNN is trained on the spinless uncompensated system with $L\approx228.6$\AA, apart from the system with $L\approx400.1$\AA.
The critical concentration estimated from this plot is $\approx 1.55\times 10^{18}\mathrm{cm}^{-3}$, 
which is consistent with the result of the larger system.
In Fig.\ref{fig4} we show the effective multifractal exponent $\tilde{\alpha}_{0}$ as a function of donor concentration
for three system sizes $L\approx 228.6$, $342.9$, and $400.1$ \AA.
The solid lines are the fits with second order polynomials for each system size.
The curves cross, to a good approximation, at a single point.
The transition concentration can be estimated from the crossing point of the curves for the largest and smallest systems as 
$1.55 \times 10^{18}$ cm$^{-3}$.
This is in good agreement with the prediction of the CNN.

\subsection{Spinless model of a compensated doped semiconductor}

\begin{figure}[t]
  \includegraphics[width=0.985\hsize]{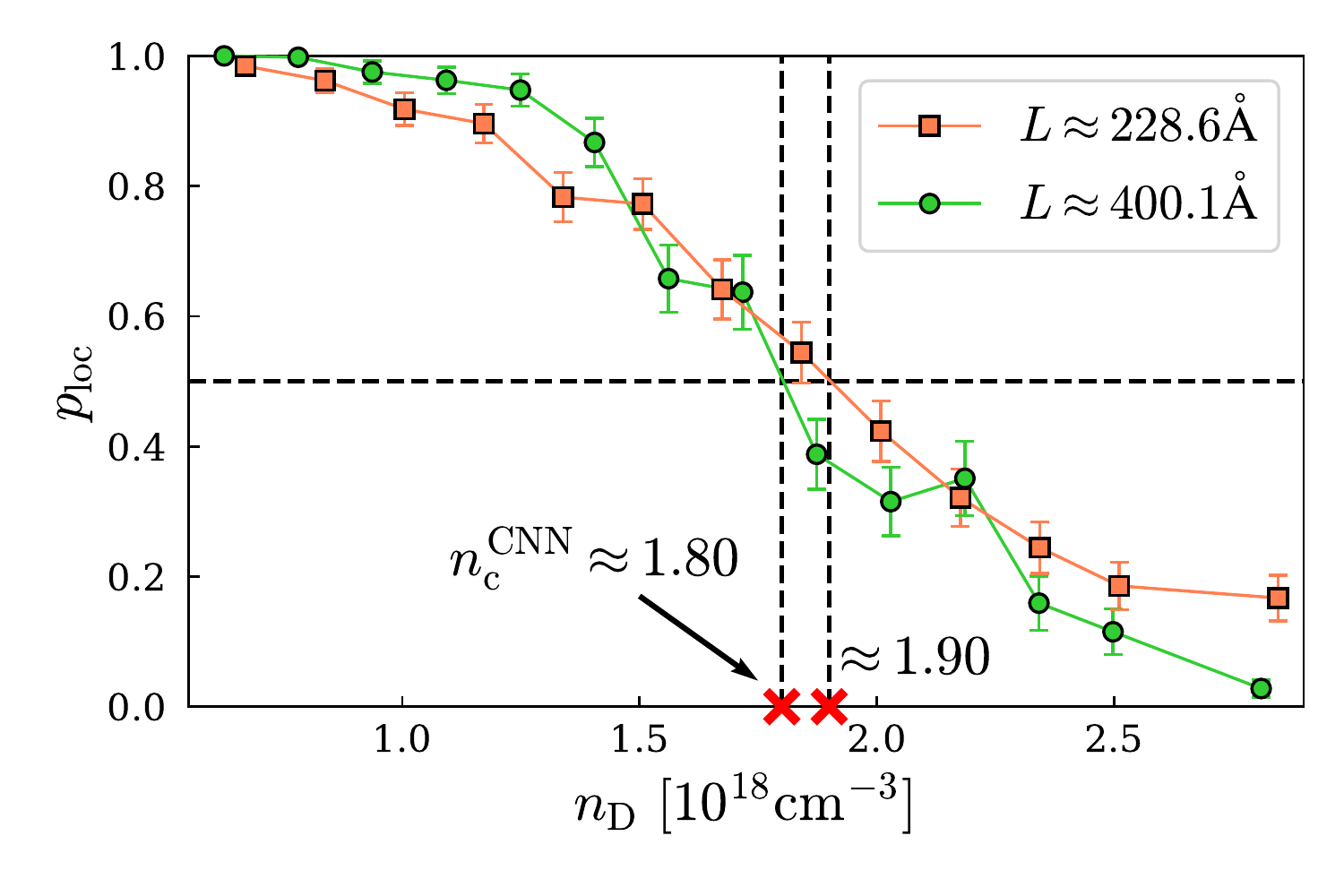}
  \caption{ 
  The probability $p_{\mathrm{loc}}$ that a Kohn-Sham eigenfunction is localised as a function of the
  doping concentration $n_{\mathrm{D}}$ for a spinless compensated doped semiconductor.
  The compensation is fixed at $50\%$ and the probability is plotted as a function of the
  concentration of donor impurities $n_{\mathrm{D}}$.
  The CNN used has been trained with data for a spinless uncompensated doped semiconductor.
  The prediction for the critical concentration is taken as the concentration where $p_{\mathrm{loc}}=0.5$.
  The estimated values are $n_\mathrm{c}\approx 1.80\times 10^{18}\mathrm{cm}^{-3}$ for $L\approx400.1$\AA\; and $n_\mathrm{c}\approx 1.90\times 10^{18}\mathrm{cm}^{-3}$ for $L\approx228.6$\AA.}
  \label{fig5}
\end{figure}

\begin{figure}[t]
  \includegraphics[width=\hsize]{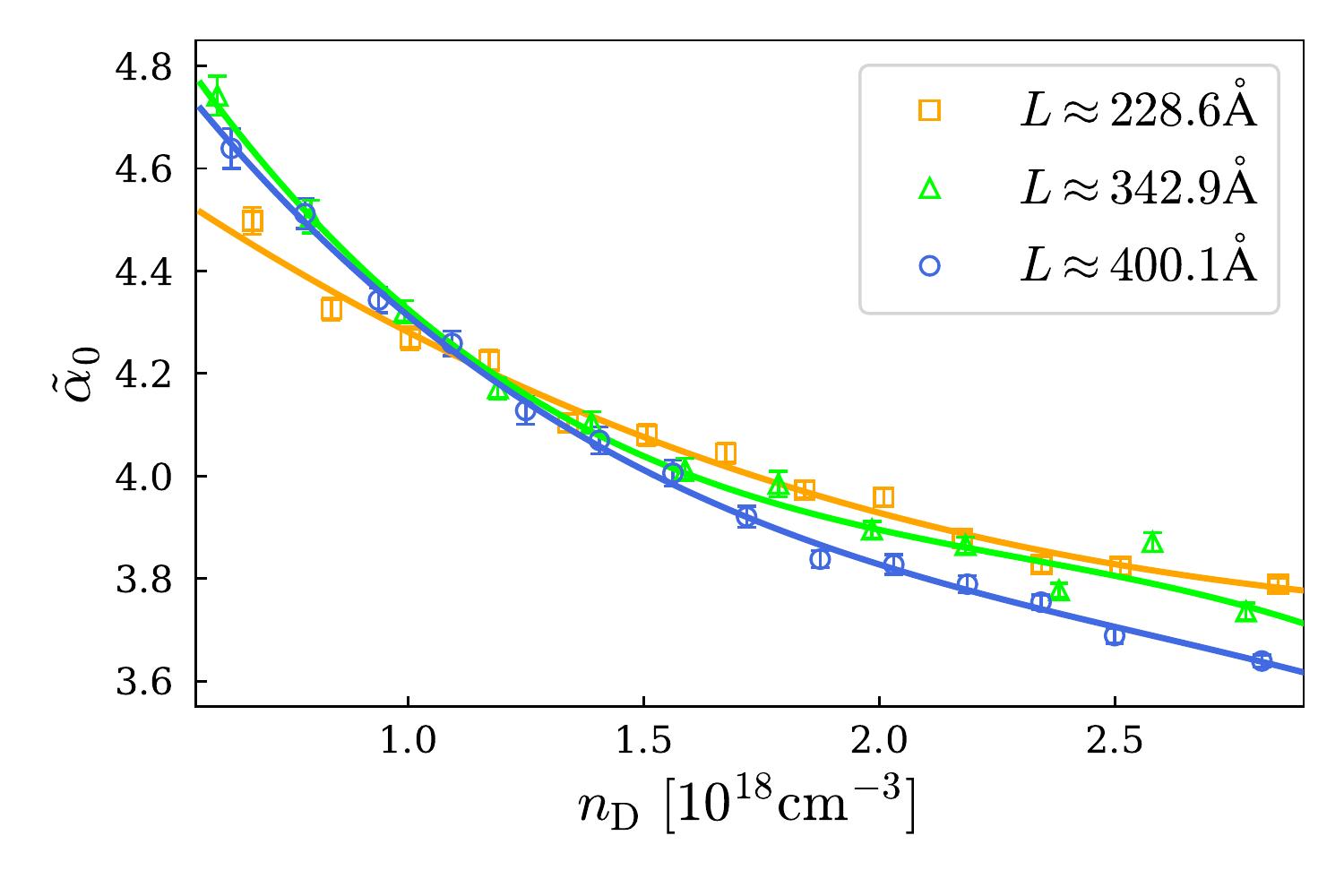}
  \caption{
  Multifractal exponent $\tilde{\alpha}_{0}$ for a spinless compensated system.
  Three system sizes were simulated: $L\approx228.6$\AA (square), $L\approx342.9$\AA (triangle), and $L\approx400.1$\AA (circle), with  
  ensembles of 103--115, 76--90, and 59--89 samples, respectively.
  No clear estimate of the critical concentration is possible:
  there is no clear common crossing point and for the
  lower concentrations the size dependence is not monotonic.}
  \label{fig6}
\end{figure}

The system now includes randomly distributed donor and acceptor ions.
The ratio of the compensation is fixed at 50\%,
\begin{equation}
    \frac{n_{\mathrm{A}}}{n_\mathrm{D}} = 0.5.
\end{equation}
We focus on the effect of compensation and neglect the spin degree of freedom.
The probability reported by the CNN trained with data for the
spinless uncompensated system is 
plotted in Fig.\ref{fig5} as a function of the donor 
concentration for two system sizes $L\approx228.6\AA$ and $L\approx400.1\AA$.
The number of samples at each data point is 103 and 59, respectively.
The data for both system sizes are in reasonable agreement and we conclude that
the CNN prediction for the critical concentration is approximately
$n_{\mathrm{c}} \approx$ 1.80 $\times 10^{18}$ [cm$^{-3}$].
The multifractal analysis is shown in Fig.\ref{fig6}.
We performed the calculations for three system sizes $L\approx400.1$\AA, $L\approx342.9$\AA, and $L\approx228.6$\AA.
The fits with third order polynomials are also shown in the figure.
For lower concentration $n_{\mathrm{D}}\lessapprox1.2\times10^{18}\mathrm{cm}^{-3}$, 
the system size dependence of $\tilde{\alpha}_{0}$ is not monotonic in
contradiction with the behaviour expected on the basis of  finite size scaling.
Furthermore, the crossing point is ambiguous and is not uniquely determined.
Thus, in this case, we are unable to predict the critical concentration on the basis of
the multifractal analysis and, therefore, we cannot verify the CNN prediction independently.

\subsection{Effect of spin and compensation on $n_{\mathrm{c}}$}

In this section, we discuss the origin of the change in $n_{\mathrm{c}}$  seen in Table~\ref{table:result} 
when either spin or compensation is included.

For the spinless uncompensated case [model (A)] the impurity band is fully occupied, unless it merges with the conduction band.
In this case delocalisation of the Kohn-Sham orbitals can only occur after the impurity band and 
the conduction bands have merged.
The situation is quite different for models (B) and (C), since
in both cases we expect the impurity band to be half filled.

\begin{figure}[H]
  \includegraphics[width=\hsize]{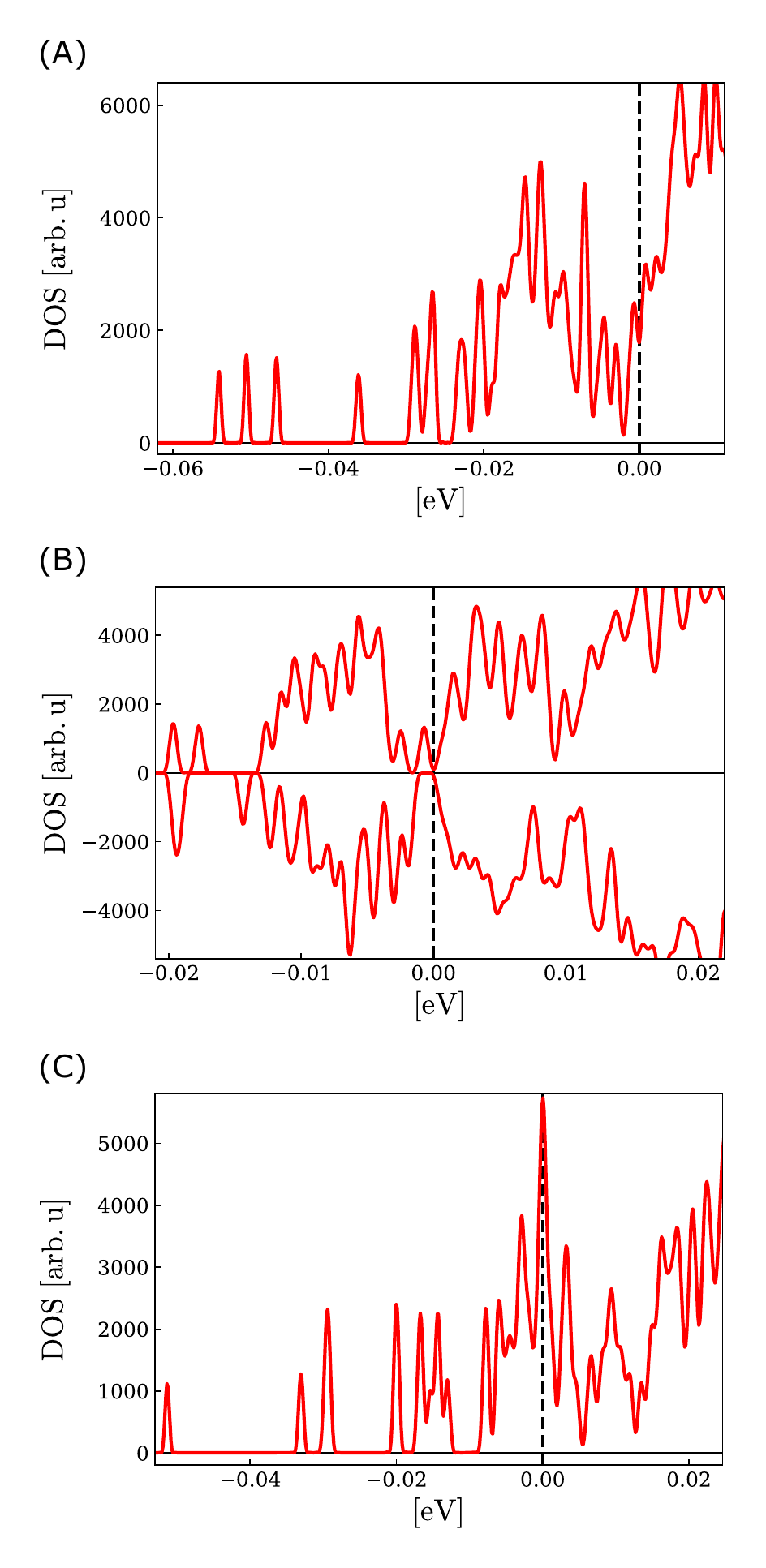}
  \caption{Density of states for the spinless model of an uncompensated semiconductor (A), the model of an uncompensated semiconductor including spin (B), and the spinless model of a compensated semiconductor (C).
    The donor concentration is fixed at $0.937 \times 10^{18} \ \mathrm{cm}^{-3}$,
    which corresponds to the insulating phase.
    Energies are shown relative to the Fermi level (vertical dashed line).
    In the panel (B), positive values correspond to the spin-up component and negative values to the spin-down component.}
  \label{fig7}
\end{figure}

In Figures \ref{fig7} (A), (B), and (C), we plot the density of states for a sample for each model.
The donor concentration has been set to the same value, which corresponds to the
insulating regime in all the three models.
The density of states were calculated using the kernel polynomial method.~\cite{Weisse2006}
The Fermi level is indicated by a vertical dashed line.

For the model with spin, a gap is observed at the Fermi level. 
Since the density of states is suppressed around this gap, the Kohn-Sham wavefunctions are 
strongly localised.
A sufficient concentration of electrons is then needed to overcome this enhancement of Anderson localisation.
The mechanism leading to the gap is not clear to us. 
The gap is absent in both the spinless simulations suggesting the importance of spin.
So an explanation in terms of the Efros-Shklovskii Coulomb gap~\cite{Efros1975}, in which spin plays no role,
seems difficult.

For both the spinless uncompensated and compensated models [models (A) and (C)] the impurity band and the conduction band are not clearly
separated and a direct comparison of the critical concentrations seems reasonable.
The critical concentration of the compensated case is larger than the uncompensated case.
A plausible explanation for compensation leading to an increase of the critical concentration 
can be found by quantitatively estimating
the randomness of an effective tight-binding Hamiltonian $H$ as follows.

The atomic orbital of an isolated impurity is the hydrogenic $1s$ orbital with appropriate
scaling for the effective mass and relative dielectric constant.
We estimate matrix elements $H_{IJ}$ of the effective Hamiltonian $H$  between atomic sites $I$ and $J$ as follows:
\begin{equation}
    H_{IJ} \equiv \int d^{3}r \; \phi_{1s}(\vec{r}-\vec{R}_{I})\; \mathcal{H}_{\mathrm{eff}}\; \phi_{1s}(\vec{r}-\vec{R}_{J})\,.
  \label{eq:hij}
\end{equation}
Here, 
\begin{equation}
    \mathcal{H}_{\mathrm{eff}} = -\dfrac{1}{2m^{*}}\nabla^{2} + V_{\mathrm{eff}},
\end{equation}
is the effective Hamiltonian, 
and
\begin{equation}
    \phi_{1s}(\vec{r}) = \dfrac{1}{\pi^{\frac{1}{2}}}\dfrac{1}{(a_{\mathrm{B}}^{*})^{\frac{3}{2}}}\exp\left(-\dfrac{r}{a_{\mathrm{B}}^{*}}\right),
\end{equation}
is the $1s$ orbital, $a_{\mathrm{B}}^{*}$ is the effective Bohr radius defined as $(\varepsilon_{\mathrm{r}}/m^{*}) a_{\mathrm{B}}$.
In this analysis for the spinless cases, the superscripts representing the spin degrees of freedom are omitted.
We expect that $H_{II}$ will correspond roughly with the diagonal term in Anderson's model of localisation, and $H_{IJ}$ for neighbouring atomic sites $I$ and $J$ with the hopping term.
The standard deviation of $H_{IJ}$ then represents the magnitude of the randomness and the mean value of $H_{IJ}$ for $I \neq J$ represents the hopping intensity.
The transport properties are determined by the ratio of the randomness to the hopping intensity.
We estimate the mean value of the hopping intensity $\mu_{t}$ and the standard deviations of the diagonal term $\sigma_{\epsilon}$ and the hopping intensity $\sigma_{t}$ for the compensated semiconductor as shown in Table~\ref{table:result} at a donor concentration of $0.937 \times 10^{18} \ \mathrm{cm}^{-3}$.
We argue that, for a given donor concentration, the model of a compensated semiconductor has greater randomness of the diagonal term $\sigma_{\epsilon}$ while their hopping properties $\mu_{t}$ and $\sigma_{t}$ are similar.
Therefore, the critical concentration should be larger because of the stronger randomness of the diagonal term.
The increase of $\sigma_{\epsilon}$ (but not $\mu_{t}$ and $\sigma_{t}$) due to compensation
is beyond the scope of this paper, which is an interesting problem left for the future.

\section{Conclusion}

In this paper, we investigated the generalisation capability of a CNN 
to determine the critical concentration of the metal-insulator transition
in a model of a doped semiconductor.

A CNN trained with Kohn-Sham eigenfunctions from DFT calculations in a 
spinless model of an uncompensated doped semiconductor
assesses correctly eigenfunctions from a model of a doped semiconductor with spin.
The predicted critical concentration is in good agreement with an independent 
prediction using multifractal analysis and finite size scaling.

We also used the same CNN to analyse eigenfunctions from a spinless model of a 
compensated doped semiconductor.
The CNN gives a clear prediction for the critical concentration but, in this case,
the multifractal analysis failed and the CNN estimate could not be independently verified.

The fact that CNN trained with eigenfunctions from a spinless model of
a doped semiconductor can be applied to assess eigenfunctions for a model of a doped 
semiconductor that includes spin may be useful.
The self-consistent calculations involved in finding the Kohn-Sham eigenfunctions
in models with spin are considerably more demanding since the spin density
must also be optimised.
Since CNNs require large training data sets to be useful, this means that 
considerable time could potentially be saved by training with a spinless model.

Before concluding this paper, we note that
the uncompensated model with spin [model (B)] has the gap at the Fermi level [Fig.~\ref{fig7}(b)] and
the spinless compensated model [model (C)]
has larger variance of the diagonal components $\sigma_{\epsilon}$ (Table~\ref{table:result}),
which explain why the critical concentrations are higher
than the case of the spinless uncompensated model [model (A)].
The origins of the gap in model (B) and the larger variance in model (C) remain
unclear.
Detailed analysis for the spin density, including its real space distribution, may be helpful 
to understand the physics of model (B).
The difference between models (A) and (C) is that 
some of negative charges are bounded to the acceptor ions in the latter, while electronic neutrality is preserved in both models.
Analysis for the difference in the electronic charge density between models (A) and (C), especially fluctuation of the charge density, may explain the difference in the variance.

\begin{acknowledgments}
This work was partly supported by JSPS KAKENHI Grant Nos. JP17K18763,  16H06345, and 19H00658.
The computation was partly conducted using the facilities of the Supercomputer Centre, the Institute for Solid State Physics, the University of Tokyo.
\end{acknowledgments}

\bibliography{References}
\end{document}